# Monolithically Integrated Ultra-Low Noise Balanced Receiver for CV-QKD

Dinka Milovančev, and Nemanja Vokić
AIT Austrian Institute of Technology, Center for Digital Safety & Security / Security & Communication Technologies,
1210 Vienna, Austria
dinka.milovancev@ait.ac.at

*Abstract* — In this work we explore monolithic opto-electronic integration platform for significant down-scaling of input-referred noise in custom designed low-noise analog front-end used for balanced photodetection. The performance of such monolithically integrated approach is compared to heterogeneously integrated solution designed in the same technology which requires wire-bonding to the balanced photodetectors. The designed circuits are targeting bandwidth above 1 GHz. The improved noise performance is leveraged against increased secure key rates and achievable reach in continuous-variable quantum key distribution (CV-QKD).

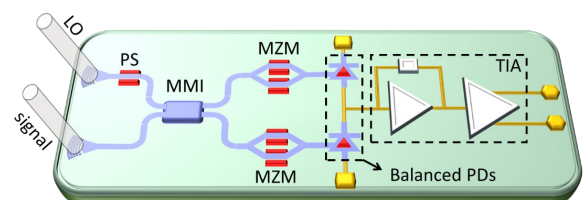

Figure 1. OEIC realization of CV-QKD receiver; PS: phase shifter for LO phase scanning, MMI: 2×2 multi-mode interferometer, MZM: Mach-Zehnder modulators, balanced PDs and a TIA

## I. Introduction

Recent progress of quantum technologies has given rise to new security treats such as quantum computing, but also new security protocols and technologies such as quantum key distribution (QKD). Continuous-variable quantum key distribution (CV-QKD) is an attractive flavor of QKD since it operates at standard telecom wavelengths, thus building on existing optical fiber infrastructure and existing pathways toward realizing necessary components on a single chip. In contrast to CV-QKD, discrete variable QKD (DV-QKD) is more mature and offers simple postprocessing but requires single-photon avalanche photodetectors (SPADs) which are more difficult to integrate and have low detection efficiencies at telecom wavelengths as well as limited speed due to the SPAD dead-time limitations. High-performance single photon detectors at 1550 nm such as superconducting nanowire single-photon detectors (SNSPD) require very low operating temperature [1] which would make their operation in elaborate QKD systems-on chip (SoC) rather challenging due to heating issues. The speed limitations make DV-QKD attractive for low-data rate long reach application scenarios [2]-[4], whereas CV-QKD can fully reach its potential for higher data rates at short and medium reach scenarios, e.g., in datacenters [5].

The critical component for CV-QKD system is the optical receiver which determines the noise and speed performance thus directly impacting available reach and QKD rate. The direct adoption of telecom grade high-speed coherent receivers for CV-QKD is not possible since these components operate at rather high frequencies (>10 GHz) which inevitably bring higher noise levels due to their wideband operation. Minimizing the electrical noise at the receiver side is of paramount importance since the detection noise, $\xi_{det}$, mainly originating from transimpedance amplifier (TIA), contributes as much as 60% of total excess noise $\xi_{tot}$ [6]. This makes receiver optimization through suitable technology choice and circuit design considerations a sweet point for yielding performance gains for CV-QKD systems. For this reason, we are exploring a monolithically integrated balanced homodyne receiver which minimizes the footprint and at the same time improves bandwidth and noise performance by implementing photonic and electronic components on the same chip as illustrated in Fig. 1. Opto-electronic integrated circuit (OEIC) development in a technology of IHP which combines silicon photonic modules with 0.25 μm BiCMOS core enables such monolithic CV-QKD receiver. At its simplest form, CV-QKD receiver with a 2×2 multimode interferometer (MMI) as an optical mixer with balanced photodiodes (PDs) fed to a TIA can be used for quantum homodyne measurement or quantum heterodyne measurement with the aid of the digital signal processing (DSP).

## II. State-of-the-Art Receivers for CV-QKD

Balanced homodyne detectors combined with low-noise TIAs are the main part of CV-QKD receivers. Differential connections between two matched photodiodes in which the anode of one photodiode is connected to a cathode of another photodiode enables cancelation of high dc photocurrent, which would otherwise enter the sensitive input TIA node. The high dc photocurrent originates from photomixing high power from the local oscillator (LO) laser with the weak quantum signal. Only the ac

This work was partly funded by the Austrian Research and Promotion Agency (FFG) through project QITTY (grant no.45004305)

component of the quantum signal mixed with the LO should be amplified by the TIA. The main requirements of CV-QKD receivers are:

- Quantum-to-classical noise ratio (QCNR) – the ratio of the LO shot noise to the noise of electrical circuitry.
- Common-mode rejection ratio (CMRR), which depends on the photodiode matching and is defined as the ratio of the difference to the sum of two photodiode currents.
- –3 dB bandwidth, which is essential for high dynamic CMRR and high QKD key rates.

For CV-QKD applications it is desirable that receivers achieve at least 10 dB of QCNR and CMRR > 30 dB [7]. Commercially available packaged components usually operate up to 1 GHz, however these usually use bulk optical components and discrete electronics leaving a lot of room for noise performance improvement, see Table I.

To improve the noise performance, some solutions in the literature make use of an operational amplifier TIA design built from discrete electronics [14], [15]. However, these solutions are usually limited to bandwidths up to 300 MHz [16] and their noise performance evidenced through classical-to-quantum noise clearance is poorer that solutions using chip-based components [4].

In the past years, an increased effort is made to scale down the CV-QKD hardware. The momentum is gained in photonic integrated circuit (PIC)-based solutions which can realize photonic components on chip, thus eliminating fiber bending loss, time skew due to different fiber lengths, and photodiode mismatch. The higher level of PD matching together with precise balancing functions through, e.g., Mach-Zehnder modulators (MZMs) result in high CMRR. Initially only the photonic parts were integrated while electrical front-end was still relying on discrete componentry thus limiting the bandwidth [17], [18]. A PIC wire-bonded to a commercial TIA chip finally enabled GHz bandwidth [19], improved noise performance and increased CMRR due to the balancing functions realized on PIC.

Recently a dedicated noise-optimized TIA designs have been reported which are suitable for wire-bonding to PICs targeting CV-QKD performance. In [20] a custom-made TIA in GaAs 100 nm pHEMT technology was wire-bonded to a silicon PIC achieving 1.5 GHz of bandwidth, with more than 20 dB quantum to classical clearance, and CMRR larger than 40 dB at 1.5 GHz. The input-referred root mean-square noise current $i_{in,rms}$ was 140 nA, which would give the noise equivalent power (NEP) of 3.6 pW/√Hz (assuming a photodiode responsivity of 1 A/W), which is a very low value for 1.5 GHz bandwidth compared to commercial products. Even lower noise performance was achieved in [21] using 55 nm CMOS technology, where designed TIA had $i_{in,rms}$ of 110 nA over 1.5 GHz of bandwidth.

TABLE I.  COMMERCIAL BALANCED DETECTORS

|  | Datasheet performance metrics | | |
| --- | --- | --- | --- |
|  | BW [MHz] | NEP [pW/√Hz] | CMRR [dB] |
| C12668-04 [8] | 400 | 30 | 30 |
| PDB425C [9] | 75 | 5.2 | 35 |
| BPD-1 [10] | 300 | 5 | 25 |
| 1817-FC [11] | 80 | 3.5 | 25 |
| PD100B [12] | 100 | 8.9 | 35 |
| WL-BPD1GA [13] | 1000 | 20 | 30 |

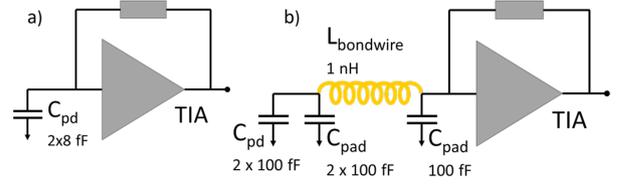

Figure 2. Equivalent circuits of simulated interface between balanced photodetectors and shunt-feedback TIA in case of a) monolithic integration and b) heterogeneous integration with external optical components.

The ultimate roadblock for even further scaling of the noise are the capacitances of the bonding pads between TIA and PIC. The bond pad capacitance amounts to ≈ 100 fF, whereas the photodiode junction capacitance itself can be in the order of 10 fF. Since $i_{in,rms}$ scales with total capacitance as $\sqrt{C_{total}}$ [22], a significant reduction is possible via monolithic integration, or alternatively a 3D integration of photonic and electronic wafers. In the next section we explore the monolithic integration design.

### III. DESIGN OF MONOLITHICALLY INTEGRATED CV-QKD RECEIVER

For the TIA design two scenarios were investigated, one for monolithic integration which is the primary focus of this work and one for heterogeneous integration which requires wire bonding to the photonic part. The latter is used for comparison in performance. Figure 2 shows the respective interfaces that were used during the design optimization. The common-emitter (CE) shunt-shunt feedback TIA topology was adopted due to its good noise performance.

#### A. TIA design

As stated, we adopted the common-emitter shunt-feedback TIA topology, see Fig. 3. To boost the gain of the forward amplifier and reduce Miller capacitance cascoded CE stage serving as a voltage amplifier is used. The cascading also helps alleviate collector emitter breakdown voltage which is 1.7 V in used technology. Additional current path through resistor $R_{C1}$ is added to increase the transconductance of the input transistor $Q_{IN}$ thus creating a gm-enhanced cascode amplifier. The additional current path also allows for the load transistor $R_C$ to be increased to boost the gain. Two emitter followers serve to decouple the high resistance output of the TIA seen at collector of Qcas toward subsequent amplifiers that would create capacitive loading, instead the output is buffered through the first emitter follower $EF_1$. The second emitter follower $EF_2$, allows to close the feedback loop without loading the TIA output with capacitance of subsequent stages and it shifts the output voltage dc level by $V_{BE}$ making it more compatible with subsequent stages.

The input-referred noise current power spectral density PSD for a shunt-feedback TIA with bipolar input transistor is [22]:

$$I_{n,TIA}^2 = \underbrace{\frac{4kT}{R_F}}_{\substack{feedback \\ resistor \\ noise}} + \underbrace{\frac{2qI_C}{\beta}}_{\substack{base\ current \\ shot\ noise}} + \underbrace{2qI_C \frac{(2\pi C_T)^2}{g_m^2} f^2}_{\substack{input-referred \\ collector\ shot\ noise}}$$

$$+ \underbrace{4kTr_b(2\pi C_D)^2 f^2}_{\substack{input-referred\ base \\ resistance\ noise\ current}}$$

(1)

Where $R_F$ is the feedback resistor, $C_T$ is total input capacitance, $C_D$ is the photodiode capacitance, $I_C$ the collector current, $\beta$ the dc current gain, $r_b$ the base resistance, and gm is the transconductance of the input transistor. The $C_T/g_m$ term scales with the third power of circuit bandwidth, and it is also connected through transit frequency $f_T$ which is technology-defined parameter. In used technology, $f_T$ of npn bipolar transistors is 210 GHz which favors a low-noise design. Additionally, the input bipolar transistor needs to be sized and biased in a way to achieve the minimum input-referred noise current. For the optimum noise performance, the total input transistor capacitances need to match the photodiode capacitance [22], at the same time the optimum collector current needs to be found which is complicated by the fact that collector and base current are not independent from each other. Therefore, the optimum sizing and biasing of input transistor for different interfaces and target bandwidth was found through parametric simulation sweeps.

The designed bandwidth was targeted at 1.5 GHz to account for possible postlayout parasitics and to aim for the performance of state-of-the-art integrated balanced receivers [20], [21]. Additionally, designs at two additional bandwidths of 5 GHz and 10 GHz were made for comparison. The input-referred noise current mean root spectral densities (RSD), the integrated $i_{in,rms}$ noise currents and −3 dB bandwidths are compared in Table II for different TIA interfaces and bandwidths. At the target bandwidth of 1.5 GHz, it can be seen that $i_{in,rms}$ has decreased from 110 nA to 32 nA when monolithic integration is used instead of heterogeneous, which would need to be wire-bonded to balanced photodetectors. Therefore, monolithic integration results in approximately 70% of noise reduction. The resulting $i_{in,rms}$ of 110 nA over 1.5 GHz of the TIA that is suitable for external wire-bonding is comparable to the state-of-the-art integrated balanced receivers for quantum communications. In [21], $i_{in,rms}$ of 110 nA over 1.4 GHz bandwidth was achieved using 55 nm CMOS technology, whereas in [20] $i_{in,rms}$ of 140 nA over 1.5 GHz was achieved in 100 nm GaAs pHEMT technology. Therefore, regardless of the used integration platform for IC design, a significant noise reduction (more than 70%) is possible with monolithic integration of balanced photodetectors with the noise-optimized TIA. Additional designs at higher bandwidths show an interesting result – by a monolithic integration the bandwidth can be extended up to 5 GHz while achieving the noise performance of a TIA designed for 1.5 GHz with external photodiodes.

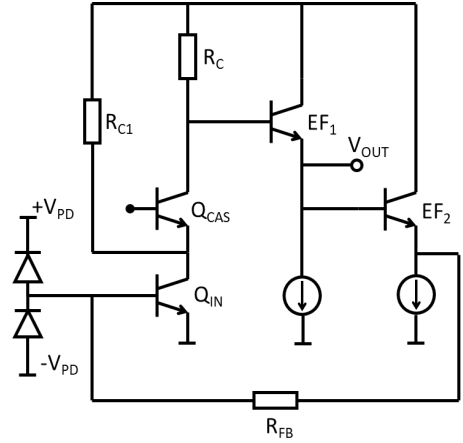

Figure 3. BiCMOS TIA schematic connected to balanced photodiodes.

TABLE II. DESIGNED TIA COMPARISON

|  | Simulation results | | |
|---|---|---|---|
|  | **BW** | **Mean RSD** | **$i_{in,rms}$** |
| Monolithic | 1.48 GHz | 0.89 pA/√Hz | 31.9 nA |
| Monolithic | 4.95 GHz | 1.85 pA/√Hz | 109 nA |
| Monolithic | 10.05 GHz | 3.2 pA/√Hz | 262 nA |
| Heterogeneous | 1.5 GHz | 3.5 pA/√Hz | 110.5 nA |
| Heterogeneous | 5 GHz | 7.05 pA/√Hz | 493 nA |
| Heterogeneous | 10.6 GHz | 19.68 pA/√Hz | 790 nA |

### B. Postlayout simulation of monotlitic OEIC and photonic layout

The postlayout simulation was carried out for monolithically integrated receiver targeting 1.5 GHz bandwidth. The transimpedance was 48.64 kΩ, the −3 dB bandwidth has slightly dropped to a value of 1.36 GHz with $i_{in,rms}$=30.4 nA. The TIA output was fed to a single-stage 50 Ω output buffer resulting in the circuit bandwidth of 1.34 GHz and $i_{in,rms}$ = 30.4 nA. The resulting transimpedance was 29.3 kΩ. The flat ac response is shown in Fig. 4, whereas the input-referred RSD is plotted in Fig. 5.

Photonic components necessary for realization of simple CV-QKD receiver are available in the process design kit optimized for 1550 nm. Two grating couplers serve for light coupling onto the PIC, one for local oscillator and one for quantum signal. The signal and the LO are mixed via a 2×2 MMI and fed to the waveguide coupled photodiodes. The optical attenuators before photodiodes serve for balancing the optical powers at the outputs of the 2×2 MMI to achieve high CMRR. The balanced photodiodes were placed as close as possible to the TIA input transistor to reduce any additional parasitics from metal connections, see Fig. 6

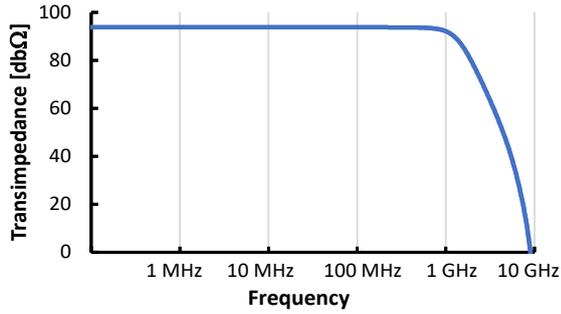

Figure 4. Simulated frequency response of the monolithically integrated CV-QKD receiver.

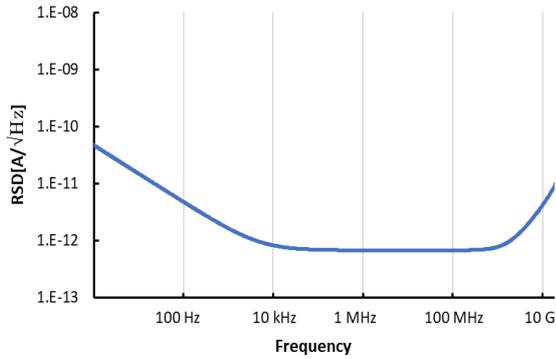

Figure 5. RSD of monolithically integrated CV-QKD receiver

## IV. CV-QKD Performance Estimation

CV-QKD performance evaluation is simulated based on theory presented in [6]. The secure key rate depends on modulation variance $V_{mod}$, the transmittance, the reconciliation efficiency, and excess noise. Modulation variance $V_{mod}$ is a variance of Gaussian probability distribution of prepared coherent state, it is directly related to the mean photon number per symbol and can therefore be tailored (by level of attenuation at the transmitter side) to the given security requirements. Channel transmittance (T) is directly influenced by channel loss, which in case of propagation through standard single mode fiber is 0.23 dB/km. The reconciliation efficiency β depends on the mutual information between transmit (Alice) and receive side (Bob) and the rate of bits that can be extracted from the information reconciliation step. The only parameter influenced by the CV-QKD receiver design is the excess noise $\xi_{tot.}$ in which the detection noise $\xi_{det}$ plays an important role. The detection noise is directly related to quantum to classical clearance, plotted in Fig. 7 based on the TIA noise simulations and supposed LO power of $P_{LO} = 10$ mW. The responsivity of the photodiodes available in the used technology is 0.7 A/W since these photodiodes are tailored for high speed and not high responsivity. As expected, the clearance is higher for lower TIA mean RDS values. The monolithically integrated TIA with 1.5 GHz bandwidth has the clearance of 29 dB at $P_{LO} = 10$ mW. If the photodiodes would be tailored for high responsivity thus reaching typical values of 1 A/W then the clearance would amount 32.08 dB.

For the simulation of receiver performance for CV-QKD, an untrusted receiver scenario is assumed in the quantum homodyne measurement. The channel loss is 0.23 dB/km, and the total detection loss includes 3 dB loss of the grating couplers and 1.55 dB losses due to the photodiode responsivity (R = 0.7 A/W). The case for optimized photonic part with edge couplers used for optical coupling with typical loss of 1 dB and high photodiode responsivity is also considered. In the model, for simplicity we suppose no quantization noise due to analog to digital converters, and no CMRR noise due to the on-chip balancing functionality. The reconciliation efficiency β of 0.96 was assumed, $V_{mod} = 6$ SNU and symbol rate of 250 Mbaud. Figure 8 reports the expected SKR for different values of channel noise ξ referred to the receiver input which may originate from different sources such as imperfect modulation, phase fluctuations, etc. Figure 8a) shows that 10 Mbit/s SKR below 10 km can be reached for moderate values of ξ. Reach of 20 km would be possible at 1 Mbit/s, whereas at ultra-short reach more than 30 Mbit/s of SKR would be possible. Figure 8b) shows significant improvements in both SKR and reach if photonic components would be optimized for best responsivity and coupling efficiency, and a much higher tolerance on channel noise. In the case of improved photonic components, the SKR could reach up to 100 Mbit/s at ultra-low reach, whereas at 10 km and ξ between 0.02 and 0.04 more than 30 Mbit/s SKR would be possible, and the longest reach would be 27 km at 1 Mbit/s SKR with ξ = 0.02.

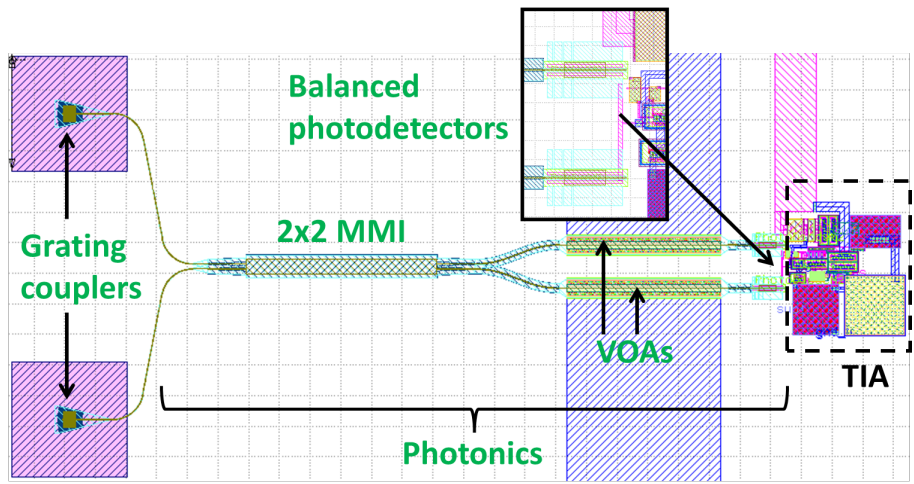

Figure 6. CV-QKD receiver layout showing the photonic and electronic parts.
MMI = multimode interferometer; TIA = transimpedance amplifier; VOA = variable optical attenuator.

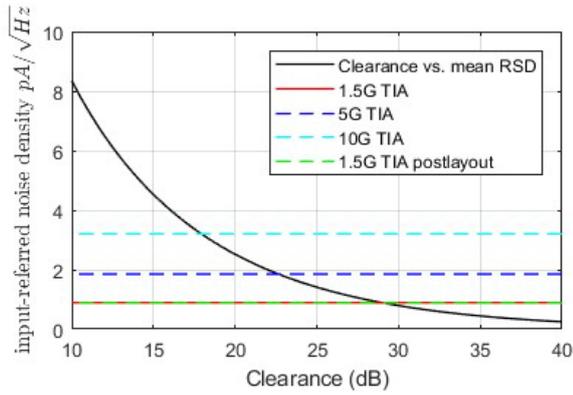

Figure 7. Clearance values based on mean RDS values.

## V. Conclusion

We have shown significant improvements in terms of low-noise operation for CV-QKD receiver design which can enable high levels of quantum to classical clearance. Compared to wire-bonded counterparts, monolithically integrated CV-QKD receiver can achieve up to 70% lower noise, which is crucial for CV-QKD operation. The silicon platform can lead to a future low-cost and small footprint CV-QKD systems. Technologically possible improvements in photonic component design are needed to fully unlock the potential of the monolithic solutions thus yielding long transmission distances with high SKR.

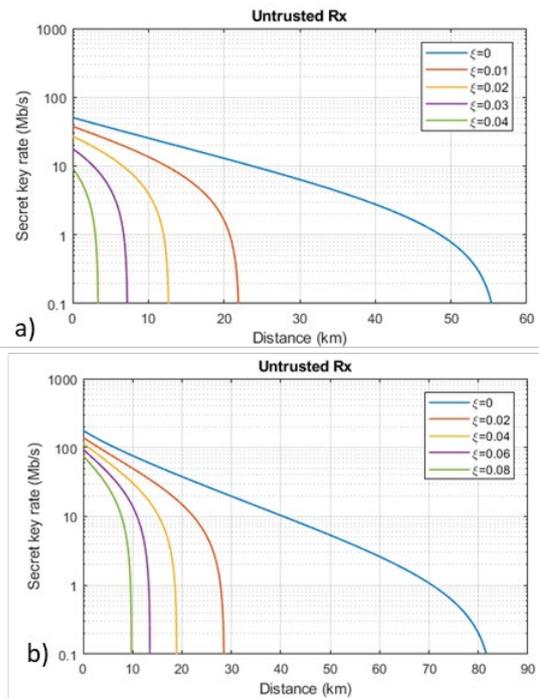

Figure 8. Simulated SKR for CV-QKD in an untrusted receiver scenario based on TIA noise performance and a) typical performance of used photonic building blocks-grating couplers with 3 dB coupling loss and photodiode responsivity of 0.7 A/W, and b) potentially optimized photonic components suitable for edge coupling with 1 dB of loss and R = 1 A/W